\documentclass{article}

\usepackage{arxiv}

\usepackage[utf8]{inputenc} 
\usepackage[T1]{fontenc}    
\usepackage{hyperref}       
\usepackage{url}            
\usepackage{booktabs}       
\usepackage{amsfonts}       
\usepackage{nicefrac}       
\usepackage{microtype}      
\usepackage{lipsum}
\usepackage{graphicx}
\graphicspath{ {./images/} }

\title{Emergent Regulatory Response and Shift of Half induction point under Resource Competition in Genetic circuits}

\author{
 Priya Chakraborty \\
 Department of Physics\\
   National Institute of Technology,\\
Durgapur, India. \\
\texttt{pc.20ph1104@phd.nitdgp.ac.in }
   \And
 Sayantari Ghosh \\
 Department of physics\\
  National Institute of Technology,\\
  Durgapur, India\\
  \texttt{sayantari.ghosh@phy.nitdgp.ac.in} \\
}

\begin{document}
\maketitle
\begin{abstract}
Synthetic genetic circuits are implemented in living cells for their operation. During gene expression, proteins are produced from the respective genes, by formation of complexes through the process of transcription and translation. In transcription the circuit uses RNAP, etc. as resource from the host cell and in translation, ribosome, tRNA and other cellular resources are supplied to the operating circuit. As the cell contains these resources in limited number, the circuit can suffer from unprecedented resource competition which might destroy the circuit functionality, or introduce some emergent responses. In this paper, we have studied a three-gene motif under resource competition where interesting behaviour, similar to regulatory responses occur due to limited supply of necessary resources. The system of interest exhibits prominent changes in behaviour which can be observed experimentally. We focus on two specific aspects, namely, dynamic range and half-induction point, which inherently describe the circuit functionalities, and can be affected by corresponding resource affinity and availability.
\end{abstract}

\keywords{Resource sharing \and Half induction point \and Range of dynamic region}

\section{Introduction}
Gene regulatory circuits, implemented in cell are coupled with the host cell for its operation. This coupling is extremely complex and non-linear in nature. As synthetic biology is growing, from last few years researchers have started to focus on the synthetic construct along with the effect of the rest of the regulatory network on it\cite{brophy2014principles}. The network circuits couple with its host in terms of growth and alloted resource like RNA Polymerase, ATP, ribosome, tRNA for necessary gene expression. These resource-dependent coupling, most of the times, does not go without any effect.\\ 
To understand and interpret behaviour of these genetic circuits, we often model the dynamics mathematically to study their overall functionality. While modelling, the effects of concerned network motif dynamics are considered but the bigger network effects often get ignored, which might cause a failure in output prediction and the circuit response becomes inexplicable. So from last few years, scientists are concentrating to develop proper understanding of host-circuit inter-dependency which includes all the local and global aspects.
\\ Some significant recent work in this domain concern the effect of growth of the host cell, which is found to modulate the circuit response significantly. The metabolic burden it generates, puts a repression effect in cell growth while in other hand growth causing dilution imposes repression in the cell circuit, which many-a-times can act as emergent feedback, significantly affecting the circuit functionality. While the effects of growth has been well-established through several recent experimental studies\cite{klumpp2009growth}, theoretical framework for growth-related effects has also been developed \cite{tan2009emergent,ghosh2011phenotypic,ghosh2012emergent}. However, the effect of resource sharing is still being explored while a bunch of experimental studies have shown this can have an extreme impact on the circuit functionality\cite{cookson2011queueing}.\\
Two major resources for protein production are basically required in forms of RNAP and ribosomes in the step of transcription and translation \cite{thecell2000} which are supplied to the circuit of interest from the available pool of the host cell. Also, tRNAs, transcription factors, degradation machinery are other important resources, taken from the host cell. 
Now in the cell these resources are not available infinitely\cite{zabet2013effects}. The cell allocates its resources for its own housekeeping functions as well as stress response. But, at the same time, the functionality of synthetic circuits depends on it. In this situation, the genes of concerned motif can face a competition with its nearby genes, either heterologous or endogenous, which are using the same resource pool. As a result, the circuit functionality can abruptly be damaged or response can also change unexpectedly. These effects are mathematically and experimentally being explored only very recently \cite{gyorgy2015isocost,weisse2015mechanistic}, and several interesting observations are needed in this domain to make our understanding of this competition and its effects more clear.
\\In this work, we focus on this phenomena of resource competition in a three-gene motif. We know that protein production mostly follows sigmoidal growth curve where output protein concentration follows a sigmoid increase with the input signal concentration. The stiffness of the curve is known to depend upon many factors like positive or negative auto regulation \cite{verma2006biological}, noise etc. From early sixties, Hill function model is widely used to represent the ability of proteins to modulate the gene expression dynamics very accurately\cite{bottani2017hill,bose2012origins}. The prediction of sigmoid binding curve of this model is universally evidenced. Half induction point of a sigmoidal Hill curve has various physical significance and is very closely related to regulatory pathways present in a circuit. A significant impact was observed by Madar et al.\cite{madar2011negative}, when presence of a negative auto regulation is found to linearize the input-output response curve in regulatory network of the arabinose utilization system in \textit{E. coli}. This in turn also regulates the range of dynamic region, i.e., the acceptable range of input signals up to which the system can respond through output, and controls the shift of half induction point. Several theoretical and experimental studies have shown similar results\cite{madar2013design,savageau1974comparison,nevozhay2009negative,dublanche2006noise,verma2006biological}. In this work, we look for similar effects, driven by resource competition. 
While competing for resources, the genes can withdraw resource with greater affinity from the total cellular resource pool, so that others might face a scarcity of resource for their respective production. This generates a regulatory effect on the the production process even when there is no direct regulatory motif exist in the circuit. 
We discover that the resource competition can cause changes in sigmoidal response curve, similar to regulatory behaviour, where no regulation is actually present. The major contributions of the paper are: 
\begin{itemize}
    \item To explore how resource competition can shift the dynamic growth curve  in protein production.
    \item To observe the dependence of the range of dynamic region on the resource allotment.
    \item To control and modify the stiffness of the growth curve, linearizing it with limited resource supply.
    \item To observe the shift in half induction point due to difference in resource affinity.
\end{itemize}
\section{Dynamic model of resource competition between regulatory network motif}

\subsection{Resource sharing between two genes.}
\begin{figure}
    \centering
    \includegraphics[width=\linewidth]{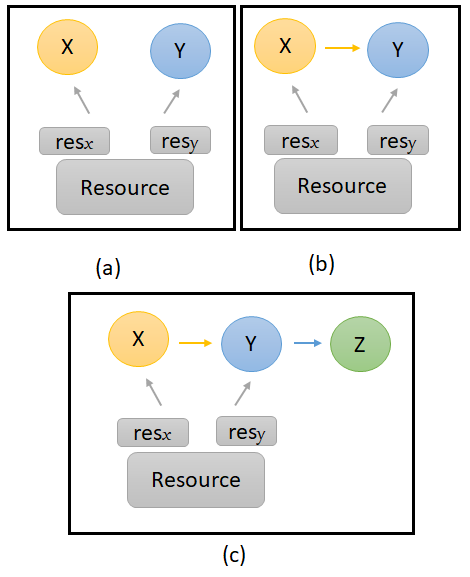}
    \caption{Model Motif. (a) $x$ and $y$ are the two proteins collecting resource from the pool with affinity $res_x$ and $res_y$. (b) $x$ and $y$ sharing the resources and also connected with a positive regulation. (c) Motif of interest where $x$ and $y$ are competing for resource, positively regulated, and $y$ also positively regulating a third protein $z$.}
    \label{fig:my_model}
\end{figure}

Let us consider a simple motif as shown in Fig. \ref{fig:my_model}(a). Let $x$ and $y$ be the two proteins, produced in the cell from their respective complex $c_x$ and $c_y$. These two are using resource, say, ribosome,  for their protein production from the same available ribosome pool $T$. This $T$ actually represents the free ribosome concentration in the immediate vicinity of the circuit interest. Let $g_x$ and $g_y$ be the mRNA copy numbers for translation process of $x$ and $y$ respectively, $\epsilon_x$ and $\epsilon_y$ are the rate of production of protein $x$ and $y$ from their corresponding complex $c_x$ and $c_y$. We focus in the process of translation here, where ribosomes bind with the mRNAs to produce the complex, and translated to protein. We further consider the resource binding affinities are not same for all genes \cite{omotajo2015distribution}; $res_x$ and $res_y$ represents the binding affinity for $x$ and $y$. On a single mRNA we have considered multiple ribosomes can bind\cite{cooper2007cell}. From the total pool $T$, as $c_x$ and $c_y$ represents the ribosome bound complex, the effective free ribosomes can be estimated by $(T-c_x-c_y)$. Here, $\delta c_x$ and $\delta c_y$ represents the complex degradation rates, while $\delta x$ and $\delta y$ are the protein degradation rates respectively for $x$ and $y$. The mathematical model represented by Eq. \ref{model initial} depicts the scenario described here, in terms of ODEs:
\begin{eqnarray}\label{model initial}
    \frac{dc_x}{dt}=res_x\;(T-c_x-c_y)\;g_x-\delta c_x\; c_x\nonumber\\
     \frac{dx}{dt}=c_x\;\epsilon_x-x\;\delta x\\
\frac{dc_y}{dt}=res_y(T-c_x-c_y)\;g_y-\delta c_y\;c_y\nonumber\\
\frac{dy}{dt}=c_y\;\epsilon_y-y\;\delta y\nonumber
\end{eqnarray}

\begin{figure*}
    \centering
    \includegraphics[width=\linewidth]{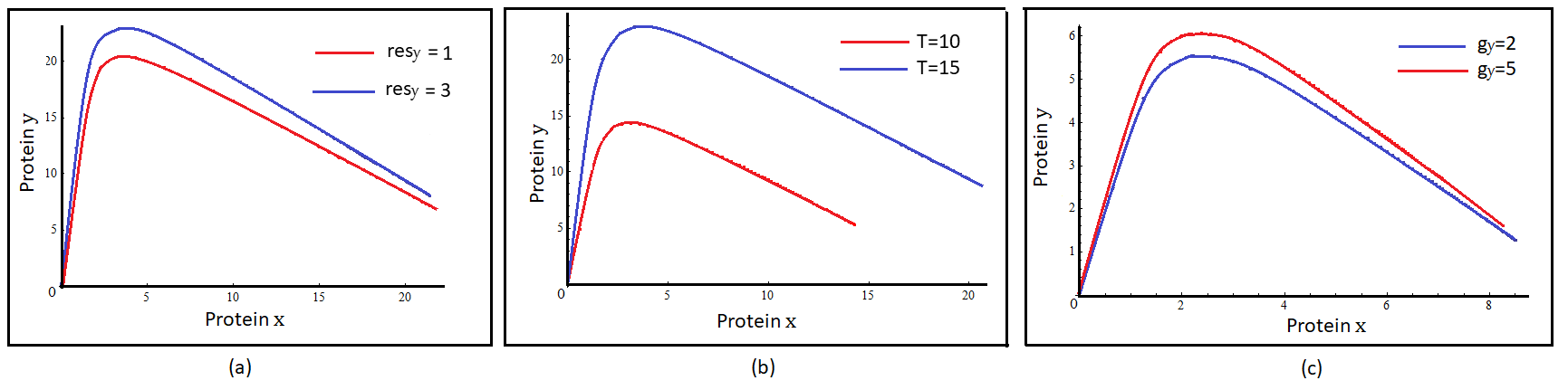}
    \caption{Production of $x$ vs. $y$ positive regulatory motif in a fixed resource budget. (a) Protein $x$ vs. protein $y$ plot. $T = 15$, $\epsilon_x=\epsilon_y=2$, $g_x = g_y = 5$, $\delta x = 
 \delta y= 1$, $n = 2$, $\delta c_x = \delta c_y = 1$. $res_x$ is changed from $0$ to $8$, red plot is for $res_y=1$ and blue plot is for $res_y=3$. (b) Protein $x$ vs. protein $y$ plot. $\epsilon_x=\epsilon_y=2$, $g_x = g_y = 5$, $\delta x = 
 \delta y= 1$, $n = 2$, $\delta c_x = \delta c_y = 1$, $res_y=3$. $res_x$ is changed from $0$ to $8$, Red curve is when $T=10$, blue curve is for $T=15$. (c) Protein $x$ vs. protein $y$ plot. $T = 10$, $\epsilon_x =\epsilon_y = 1$, $g_x = 5$, $\delta x =\delta y=1$, $n = 2$, $res_y = 3$, $\delta c_x = \delta c_y = 1$. $g_y = 2$ gives the blue plot and $g_y =5$ gives the red plot. $res_x$ is changed from $0$ to $8$.}
    \label{fig:isocost}
\end{figure*}
\subsection{Resource sharing in a positive regulatory genetic motif.}
In Fig. \ref{fig:my_model}(a), though the two genes are apparently not connected with each other, a correlation is established in terms of resource sharing in Eq. \ref{model initial}. As the total pool of resource is fixed, if one of the genes draws more ribosome for its translation with greater resource affinity, other will suffer a scarcity of resource. Keeping this competition in the equation, let us proceed one step further, and consider that $x$ and $y$ are correlated with a positive relation now, i.e., $x$ activates the production of $y$ and both collect resource from the same pool as shown in Fig. \ref{fig:my_model}(b). We have included the activation term here as represented by Eq. \ref{eqn midd} below:
\begin{eqnarray}\label{eqn midd}
    \frac{dc_x}{dt}=res_x\;(T-c_x-c_y)\;g_x-\delta c_x\;c_x\nonumber\\
\frac{dx}{dt}=c_x\;\epsilon_x-x\;\delta x\\
\frac{dc_y}{dt}=res_y(T-c_x-c_y)\;g_y-\delta c_y\;c_y\nonumber\\
\frac{dy}{dt}=\frac{c_y\;\epsilon_y\;x^n}{1+x^n}-y\;\delta y\nonumber
\end{eqnarray}

\subsubsection{Observations on Resource Competition} As $x$ positively regulates $y$, a linear response is expected between $x$ and $y$, but the limited resource plays a great role here when the same pool allots the resource for both the production. In Fig. \ref{fig:isocost}(a) from the $x$ vs. $y$ response curve of Eq. \ref{eqn midd},  we get very prominently two region, a positively regulated region followed by a competition region. While first region showing a positive correlation coming from the existing regulation $x\rightarrow y$, the region of negative correlation is very similar to the isocost lines observed in economics\cite{gyorgy2015isocost}, which clearly signifies a tight resource budget. \\
We have taken $\epsilon_x = \epsilon_y = 2$, $g_x = g_y = 5$, $\delta x=\delta y=1$, $\delta c_x =\delta c_y = 1$, for both Fig.  \ref{fig:isocost}(a) and (b), where $n$, the co-operativity, is taken as $2$ through out rest of the model results. For Fig. \ref{fig:isocost}(a), with constant resource pool $T=15$, $res_y = 1$ is giving the red curve while $res_y = 3$ gives the blue curve, when $res_x$ is continuously varied from $0$ to $6$. Here, as we plot $x$ vs.$y$, an increase in starting clearly reflects the positive regulation of $x$ to $y$, while for the greater production in a fixed budget the competition is very prominent. With an increase of $x$ for higher values, $y$ can not increase further due to its lack of resource affinity, and fall down at higher $x$ with a slope proportional with $res_y$ affinity. With a lower resource to $y$, $res_y=1$, the maximum achievable amount of $y$ is less as $y$ cannot accommodate resource for its production while with $res_y=3$ the maximum of $y$ is greater, as shown in the blue curve. 
\\For Fig. \ref{fig:isocost}(b), we keep $res_y=3$ and plot for two different values of $T$, for $15$ blue curve and for $10$, the red one. This figures can also be explained in a straight forward manner while in presence of a resource competition, greater the total resource pool, greater is the production level.
\\ In Fig. \ref{fig:isocost}(c) we get similar response as Fig. \ref{fig:isocost}(b) where we have plotted two different values of mRNA copy numbers $g_y$, when all other parameters are kept fixed at $T = 10$, $\epsilon_x =\epsilon_y = 1$, $g_x = 5$, $\delta x =\delta y=1$, $n = 2$, $res_y = 3$, $\delta c_x = \delta c_y = 1$. With a low $g_y = 2$, we get the blue curve and a comparative high $g_y =5$ gives the red plot. This indicates the study when one mRNA is present in more numbers compared to the other. Here, $res_x$ is changed for $0$ to $8$. The change in stiffness of response curve can be described by the resource competition model. Less availability of mRNA corresponding to $y$ for the blue plot ($g_y=2$) ended with a lesser production of $y$, now for the production of same concentration of $y$ as like $g_y=5$ curve $y$ collects more ribosomes from the pool which eventually decreases the resource availability to $x$ ending with a smaller value of $x$ in the competition region.
\\All these figures establish that in our model the sharing effect has genuine implications along with the positive regulation impact. 
\subsection{Input-output response of Downstream protein under Resource Competition}
Now, we consider a third downstream protein $z$ is activated by the motif model established earlier. A simple schematic model is as shown in Fig. \ref{fig:my_model}(c), where $x$ activates $y$, $y$ activates $z$ respectively. While $x$ and $y$ are mutually connected with each other in terms of resource sharing from the same resource pool $T$, let $x$ is collecting ribosomes with resource affinity $res_x$ and $y$ is collecting with affinity $res_y$ for their respective complex production $c_x$ and $c_y$. $\delta z$ is the rate of degradation of $z$ protein. The set of Eq. \ref{final equation} represents the overall model.\\
\begin{eqnarray}\label{final equation}
    \frac{dc_x}{dt}=res_x\;(T-c_x-c_y)\;g_x-\delta c_x\;c_x\nonumber\\
\frac{dx}{dt}=c_x\;\epsilon_x-x\;\delta x\\
\frac{dc_y}{dt}=res_y(T-c_x-c_y)\;g_y-\delta c_y\;c_y\nonumber\\
\frac{dy}{dt}=\frac{c_y\;\epsilon_y\;x^n}{1+x^n}-y\;\delta y\nonumber\\
\frac{dz}{dt}=\frac{y^n}{1+y^n}-z\;\delta z\nonumber
\end{eqnarray}
At equilibrium, all the rate of changes can be taken equal to zero. we study the model depicted by Eq. \ref{final equation} at steady state.
\begin{figure}
    \centering
    \includegraphics[width=\linewidth] {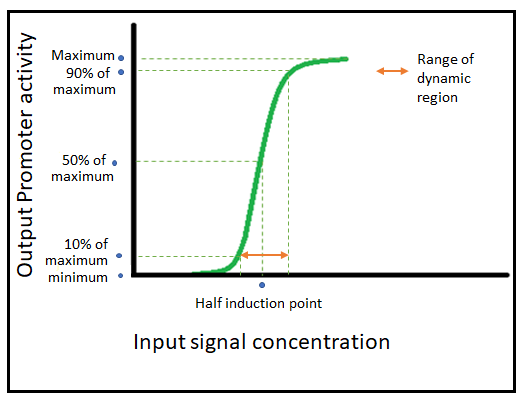}
    \caption{Schematic figure indicating the range of dynamic region and half induction point.}
    \label{schmt half ind}
\end{figure}
\section{Downstream Response emerging under Resource Competition}\label{Result section}
\begin{figure}
    \centering
    \includegraphics[width=\linewidth]{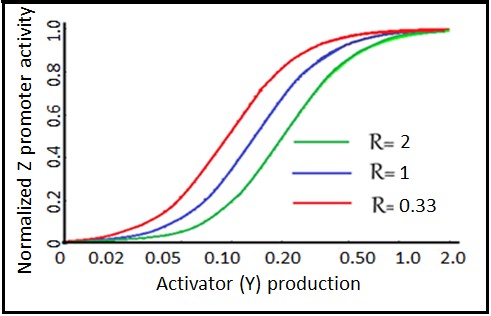}
    \caption{Shifting of dynamic region for different resource affinity ratio $R$. Activator ($y$) production vs. Normalized $z$ promoter activity plot. $T = 15$, $\epsilon_x = 2$, $g_x = g_y = 5$, $\delta x = 
\delta y = \delta z = 1$, $n = 2$, $\delta c_x =\delta c_y = 1$, $\epsilon_y$ is changed from $0$ to $2$. Green curve is for $R=2$, blue curve is for $R=1$, while $R=0.33$ represents the red one.}
    \label{fig:resource regulion}
\end{figure}
As $x$ and $y$ competes for the resources, we study how this competition effects the production of third downstream protein $z$.  From the output response Fig. \ref{fig:resource regulion}, it is clear that the production of $z$ is highly dependable on the resource sharing of $x$ and $y$. Let us elaborate our observations in this section.
\subsection{Shifting of dynamic region}
The first observation we have here is in terms of shifting of the dynamic region with resource variation. To investigate the shift in the response curve of protein $z$ followed by the sigmoidal growth, similar to Hill function model, we define the input dynamic region as the range of input signal over which the output shows $10\%$ and $90\%$  of its maximum, following existing research \cite{goldbeter1981amplified,goldbeter1984ultrasensitivity}. An illustrative figure of dynamic range is shown in Fig. \ref{schmt half ind}.
 Thus, the dynamic region is the range of the input signal over which the output response is sharply changing.  In Fig. \ref{fig:resource regulion}, with a fixed value of $res_y$ at $3$ we observe the normalised $z$ promoter activity vs. activator $y$ production rate for $3$ different values of $res_x$ when all the other parameters are same for all the plots. By the term normalised promoter activity we mean the measurement of gene product (i.e concentration of protein $z$) in a normalised fashion.
\\Here, we define a new parameter resource affinity ratio $R$ by Eq. \ref{resratio}:
\begin{equation}
    R=\frac{res_x}{res_y}
    \label{resratio}
\end{equation}
This $R$ provide us the ratio of the resource affinities, giving an estimation of allocated resources for $x$ compared to $y$, in tight resource condition. With $res_x=res_y=3$, the resource affinity ratio $R$ will be $1$, the blue sigmoid, in Fig. \ref{fig:resource regulion} shows the production of $z$. Now keeping all parameters including $res_y$ same, if we decrease resource to $x$ that is $res_x$ to $1$, the affinity ratio $R$ will be $0.33$, the curves shifts to the left (i.e., the red curve). Similarly, when $res_x$ is chosen as $6$, that is $R$ is $2$ now, the curve shifts towards right (the green sigmoid in Fig. \ref{fig:resource regulion}). \\
Here, it is important to note that Hill function co-operativity can also invariably affect the sharpness and induction level of the response curve which is not the underlying reason here.  The shift we observe here is for a constant $n=2$ and is solely the effect of resource competition upstream.\\
As all the parameters are fixed, increasing resource to $x$, that is decreasing resource to $y$, shifts the response of $z$ production, when apparently no tuning is done directly related to $z$. With less resource to $y$, the production of downstream protein $z$ is affected. A complete shifting of the dynamic region in this case is actually regulated by resource sharing while this kind of behavior were found before from auto regulatory effects\cite{madar2011negative}. With the decrease in $R$ shifts the curve left, indicating that the expression can switch on for a lower value of the $\epsilon_y$. Similarly for higher values of affinity ratio $R$, the right shift of the induction curve signifies the response switches on for a higher value of $\epsilon_y$. Thus an amplification in signal can be achieved by regulating the resource affinity.
\subsection{Shifting of half induction point}
\begin{figure}
    \centering
    \includegraphics[width=\linewidth]{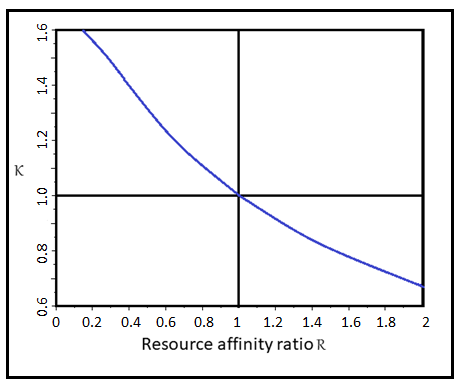}
    \caption{Shift of half induction point for increasing resource affinity ratio. $T=15$, $\epsilon_x=2$, $g_x = g_y = 5$, $\delta x = \delta y = \delta z = \delta c_x =\delta c_y=1$, $n=2$. $\epsilon_y$ is changed from $0$ to $2$.}
    \label{fig:shift half induction}
\end{figure}

The shift of dynamic region implies another important behavioral change of the output $z$ response curve in terms of shifting of half induction point. Half induction point is defined as the value of $\epsilon_y$ for which $z$ is at half of its maximum output. Fig. \ref{schmt half ind} represents a schematic estimation of half induction point. To quantify this variation, we define, half induction point ratio, $K$, in Eq. \ref{half induction}: 
\begin{equation}
K=\frac{half\; induction\; point\; when\;res_x\;=res_y}{half\; induction\; point\; for\; other\; resource\; value}
\label{half induction}
\end{equation}\\
This implies that we take the value of half induction point for same value of resource to both $x$ and $y$  and divide the half induction point value for other resource combinations with it. Now, we plot the half induction point ratio with respect to the resource affinity ratio $R$ for reference point set at $res_x = res_y = 3$. A shift in half induction point is clearly observed as shown in Fig. \ref{fig:shift half induction}. $K>1$ is for left shifting of the half induction point with respect to the half induction point for same resource, this region entirely falls on the resource affinity ratio $R<1$, which incidentally represents that for this region $x$ is getting less resource than $y$, i.e $res_x<res_y$. Similar arguments can be established for $R>1$ region. This shift in half induction point plays an important role in terms of output signal sensitivity which is completely regulated by the resource allotment to the circuit here.  
\subsection{Range of dynamic region and Response Sensitivity} 
\begin{figure}
    \centering
    \includegraphics[width=\linewidth]{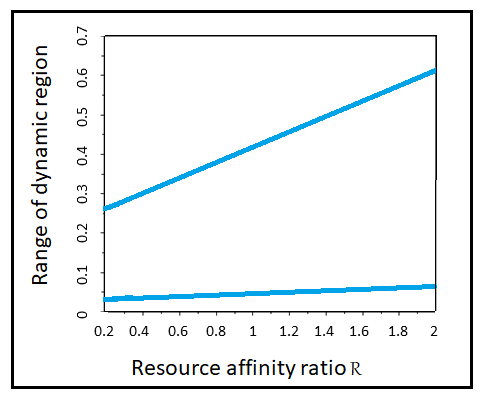}
    \caption{Range of dynamic region for increasing resource affinity ratio $R$. $T=15$, $\epsilon_x=2$, $g_x = g_y = 5$, $\delta x = \delta y = \delta z = \delta c_x =\delta c_y=1$, $n=2$, and $\epsilon_y$ is changed from $0$ to $2$.}
    \label{fig:dynamic range}
\end{figure}

Not only the shifting of the response curve but the linearization is also very prominent. Less the resource available to $y$, production of $z$ is being more linearised with a greater range of dynamic region. Our results are shown in Fig. \ref{fig:dynamic range}. Here, we define the input dynamic range in terms of the boundary values of induction level with which significant variation in response can be observed. We define the boundaries, $\epsilon_l$ and $\epsilon_u$ of the dynamics range as levels for which $z$ shows 10\% and 90\% of its maximum response $z_{max}$:
\begin{eqnarray}
    \epsilon_l=\epsilon_y\mid\{z=0.1\:z_{max}\} \nonumber\\\nonumber
    \epsilon_u=\epsilon_y\mid\{z=0.9\:z_{max}\}    
\end{eqnarray}  

The range of dynamic region has been plotted with the resource affinity ratio $R$. We observe that the range of dynamic region is increasing with the increase of $R$. Increase in $R$ means increase of $res_x$ than that of $res_y$, thus less the resource available to $y$ greater the dynamic range of $z$. This increase in dynamic range with the increase of resource affinity ratio $R$ clearly indicates the decrease in sensitivity of the output response.
\section{Discussion.}
The presence of a number of participants and their mutual regulatory linkage makes the output of an experimental genetic motif very complicated to explain. In a limited resource environment, a gene faces a competition with its neighboring genes as well as the host in general, and this can change the circuit functionality abruptly.
\\In this paper, we have explored resource sharing starting from a simple motif where two genes are competing for resource from the same pool. Greater is the resource affinity, greater is the resource allocated for its production and less is the availability for the other. Focusing primarily on the translation process where mRNA binds with the ribosomes from the available pool in order to be translated through the steps of complex formation, we have introduced the positive regulation between the genes. With the emerging isocost lines, it is evident that as long as the available resource is sufficient the circuit behavior is as expected, dominated by regulatory pathways, but for the tighter resource economy regime the limited availability of resource pushes the circuit in a competition phase. \\
In the next step, we have introduced a third positively regulated output gene $z$, and find some interesting change in behavioral output response due to resource competition. Less the availability of resource to $y$, greater is the range of dynamic region of $y$ to $z$ production curve. The linearization tendency is also very clear from the response curve, which is an impact of increase of dynamic region. There is a prominent shift in dynamic region which causes a shift also for the half induction point. An amplification in input to output signal and modification in the sensitivity of the output response can occur in the system for the unequal distribution of resource among its participants. This type of behavior is previously studied in terms of auto regulation present in the motif of interest, but the impact of resource sharing is completely unexplored yet. Our study provides a computational analysis that unequal resource affinity can cause a similar output response. Here, we wish to point out that in this study, not exact value of resource affinity but the ratio of it $R$ is the major controlling factor, which implies firm qualitative interpretations. We have also considered the local concentration of the resource and depletion framework in a local pool. This establishes the actual scenario inside the cellular environment.
\\In future, we will advance these studies to understand the underlying regulatory activities in this dynamics. Some researchers have already pointed out that the amount and direction of the half induction point shift can pinpoint the mode of regulation. Some of the well-known  multigene motifs (e.g. FFL, FBL etc.) can be studied under the resource limitation environment  and the detailed analysis of results will establish the consequence of resource competitions in protein production more accurately. Moreover, our model is also flexible enough to incorporate the transcription process in future as we know in transcription RNA polymerase behaves as the cellular resource to initiate the protein production. Thus, we are also interested to design the computational models incorporating the resource competition of all the transcriptional and translational steps for robust controlling of the synthetic genetic circuits.
\section*{Acknowledgment}
PC and SG acknowledge the support  by DST-INSPIRE, India, vide sanction Letter No. DST/INSPIRE/04/2017/002765  dated- 13.03.2019.

\bibliographystyle{unsrt}  
\bibliography{bibliography.bib}  


\end{document}